\documentstyle[12pt]{article}

\catcode`\@=11
\long\def\@makefntext#1{
\protect\noindent \hbox to 3.2pt {\hskip-.9pt  
$^{{\ninerm\@thefnmark}}$\hfil}#1\hfill}                

\def\@makefnmark{\hbox to 0pt{$^{\@thefnmark}$\hss}}  
        
\def\ps@myheadings{\let\@mkboth\@gobbletwo
\def\@oddhead{\hbox{}
\rightmark\hfil\ninerm\thepage}   
\def\@oddfoot{}\def\@evenhead{\ninerm\thepage\hfil
\leftmark\hbox{}}\def\@evenfoot{}
\def\sectionmark{}\def\subsectionmark{}}

\renewcommand{\thefootnote}{\fnsymbol{footnote}}

\renewcommand{\subsubsection}[1]
{\vspace*{0.6cm}\addtocounter{subsubsectionc}{1}
        \noindent
{\normalsize\rm\thesectionc.\thesubsectionc.\thesubsubsectionc. 
        #1}\par\vspace*{0.6cm}}

\newcounter{appendixc}
\newcounter{subappendixc}[appendixc]
\newcounter{subsubappendixc}[subappendixc]

\renewcommand{\appendix}[1] {\vspace*{0.6cm}
        \refstepcounter{appendixc}
        \setcounter{figure}{0}
        \setcounter{table}{0}
        \setcounter{equation}{0}
        \renewcommand{\thefigure}{\Alph{appendixc}.\arabic{figure}}
        \renewcommand{\thetable}{\Alph{appendixc}.\arabic{table}}
        \renewcommand{\theappendixc}{\Alph{appendixc}}
        \renewcommand{\theequation}{\Alph{appendixc}.\arabic{equation}}
        \noindent{\bf Appendix \theappendixc #1}\par\vspace*{0.4cm}}

\def\abstracts#1{{
       
\centering{\begin{minipage}{12.2truecm}\footnotesize\baselineskip=12pt\noindent
        \centerline{\footnotesize ABSTRACT}\vspace*{0.3cm}
        \parindent=0pt #1
        \end{minipage}}\par}} 

\renewenvironment{thebibliography}[1]
        {\begin{list}{\arabic{enumi}.}
        {\usecounter{enumi}\setlength{\parsep}{0pt}
\setlength{\leftmargin 1.25cm}{\rightmargin 0pt}
         \setlength{\itemsep}{0pt} \settowidth
        {\labelwidth}{#1.}\sloppy}}{\end{list}}

\newcounter{itemlistc}
\newcounter{romanlistc}
\newcounter{alphlistc}
\newcounter{arabiclistc}

\newcommand{\fcaption}[1]{
        \refstepcounter{figure}
        \setbox\@tempboxa = \hbox{\footnotesize Fig.~\thefigure. #1}
        \ifdim \wd\@tempboxa > 6in
           {\begin{center}
        \parbox{6in}{\footnotesize\baselineskip=12pt Fig.~\thefigure. #1}
            \end{center}}
        \else
             {\begin{center}
             {\footnotesize Fig.~\thefigure. #1}
              \end{center}}
        \fi}

\newcommand{\tcaption}[1]{
        \refstepcounter{table}
        \setbox\@tempboxa = \hbox{\footnotesize Table~\thetable. #1}
        \ifdim \wd\@tempboxa > 6in
           {\begin{center}
        \parbox{6in}{\footnotesize\baselineskip=12pt Table~\thetable. #1}
            \end{center}}
        \else
             {\begin{center}
             {\footnotesize Table~\thetable. #1}
              \end{center}}
        \fi}

\def\@citex[#1]#2{\if@filesw\immediate\write\@auxout
        {\string\citation{#2}}\fi
\def\@citea{}\@cite{\@for\@citeb:=#2\do
        {\@citea\def\@citea{,}\@ifundefined
        {b@\@citeb}{{\bf ?}\@warning
        {Citation `\@citeb' on page \thepage \space undefined}}
        {\csname b@\@citeb\endcsname}}}{#1}}

\newif\if@cghi
\def\cite{\@cghitrue\@ifnextchar [{\@tempswatrue
        \@citex}{\@tempswafalse\@citex[]}}
\def\citelow{\@cghifalse\@ifnextchar [{\@tempswatrue
        \@citex}{\@tempswafalse\@citex[]}}
\def\@cite#1#2{{$\null^{#1}$\if@tempswa\typeout
        {IJCGA warning: optional citation argument 
        ignored: `#2'} \fi}}

 1
 1
 1

\font\ninerm=cmr9

\def\lsim{\mathrel{\mathop{\kern 0pt <}\limits_{\displaystyle\sim}}}
\def\gsim{\mathrel{\mathop{\kern 0pt >}\limits_{\displaystyle\sim}}}
\begin{document}

\centerline{\normalsize\bf RENORMALONS AND FIXED POINTS\footnote
{Research supported in part by the EC program `Human Capital and Mobility',
Network `Physics at High Energy Colliders' ,contract CHRX-CT93-0357 (DG12 COMA).
}}
\baselineskip=22pt

\vspace{0.6cm}
\centerline{\footnotesize Georges GRUNBERG}

\baselineskip=13pt
\centerline{\footnotesize\it Centre de Physique Th\'eorique de l'Ecole
 Polytechnique\footnote{CNRS UPRA 0014}}
\baselineskip=12pt
\centerline{\footnotesize\it 91128 Palaiseau Cedex - France}
\vspace{0.6cm}
\centerline{\footnotesize E-mail: grunberg@orphee.polytechnique.fr}

\vspace{0.9cm}
\abstracts{The connection between renormalons and power corrections is 
investigated for the typical infrared renormalon integral assuming the effective
 coupling constant has an infrared fixed point of an entirely perturbative 
origin. It is shown that even then  the full answer differs from the Borel sum 
by a power correction.A comparaison with the analogue results when the fixed
 point is generated by the explicit addition of non perturbative power 
suppressed terms is given.}
\newpage 
\pagestyle{plain}
\normalsize\baselineskip=15pt
\setcounter{footnote}{0}
\renewcommand{\thefootnote}{\alph{footnote}}
It is generally believed that the large order behavior associated to infrared 
(IR) renormalons,which makes QCD perturbation theory ``non Borel summable'', 
reflects an inconsistency related to the Landau ghost. In this note, I examine 
how the renormalon problem is resolved when
the coupling constant is free of Landau singularity and approaches a non trivial
 IR 
fixed point at small momenta.Most of the paper is devoted to the case where 
 the fixed point arises entirely within a perturbative framework, through
higher 
order perturbative corrections. I will show that in this case, IR
renormalons are also present, but the exact result differs from the Borel sum 
by a (complex) power correction, which removes all inconsistencies.At the end,
some short comments will be made on the alternative case where the fixed point 
is generated by the explicit introduction of non perturbative terms.

Let us first review the standard argument\cite{{Par},{Mue}} for renormalons. 
Consider the typical IR renormalon integral :

\begin{eqnarray}
R(\alpha) = \int_{0}^{Q^{2}} n \frac{dk^{2}}{k^{2}} \left(\frac{k^{2}}{Q^{2}}
\right)^{n}
\alpha_{eff} (k/Q,\alpha) 
\end{eqnarray}
where $\alpha$ is the coupling at scale $Q$ in some arbitrary renormalization
scheme, and $\alpha_{eff}(k)$ a renormalization group (RG) invariant effective 
coupling
(I assume $n > 0$, so that the integral in Eq.~(1) is IR convergent order by 
order 
in perturbation theory). Let us compute the perturbative Borel transform $R(z)$.
 If
$R(\alpha)$ has the formal power series expansion :

$$
{R_{PT}(\alpha) = \mathop{\sum}\limits_{p=0}^{\infty} r_p\ \alpha^{p+1},}
$$
$R(z)$ is defined  by  :
\begin{equation}
R(z) = \mathop{\sum}\limits_{p=0}^{\infty}\ \frac{r_p}{p!}\ z^p 
\end{equation}
The series Eq.~(2) are believed to have a finite convergence 
radius (at the difference of those for $R(\alpha)$), and therefore
allow for an all order definition of $R(z)$. One can then {\em define} an 
all order ``perturbative'' resummed $R_{PT}(\alpha)$ by the Borel
representation:

\begin{eqnarray}
R_{PT}(\alpha) = \int_0^{\infty} dz\ exp \left(- \frac{z}{\alpha}\right) R(z) 
\end{eqnarray}
It is convenient to express $R(z)$ in term of the Borel transform of  
$\alpha_{eff}$ :

\begin{eqnarray}
\alpha_{eff}(k/Q,\alpha) = \int_0^{\infty} dz\ exp \left(- \frac{z}{\alpha}
\right)
\alpha_{eff}(k/Q,z) 
\end{eqnarray}
Inserting Eq.~(4) into Eq.~(1), and interchanging the order of
integrations, one 
indeed
recovers Eq.~(3) with $R(\alpha) = R_{PT}(\alpha)$ and :

\begin{eqnarray}
R(z) = \int_0^{Q^2} n \frac{dk^2}{k^2} \left(\frac{k^2}{Q^2}\right)^n 
\alpha_{eff}(k/Q,z) 
\end{eqnarray}
IR renormalons arise as an IR divergence \cite{Par} of the integral in Eq.~(5), 
resulting from the IR behavior
of $\alpha_{eff} (k/Q,z)$ (which follows solely from the RG invariance of 
$\alpha_{eff}$)~:

\begin{eqnarray}
\alpha_{eff} (k/Q,z) &\simeq &\alpha_{eff}(z) exp 
\left\lbrack -z \left(\beta_0\ ln \left(\frac {k^2}{Q^2}\right)
- \frac{\beta_1}{\beta_0} ln\ ln \left(\frac{Q^2}{k^2}\right)\right)\right
\rbrack\nonumber\\
 k^2 &\ll &Q^2 
\end{eqnarray}
where $\beta_0$ and $\beta_1$ are (minus) the one and two loop beta
function coefficients.
Using Eq.~(6) into Eq.~(5) one indeed finds for $z \rightarrow z_n n/\beta_0$ :

\begin{eqnarray}
R(z) \simeq \alpha_{eff} (z \simeq z_n) \frac {\Gamma(1+\delta)} {n^{\delta}}
\ \frac{1}{(1 - \frac{z}{z_n})^{1+\delta}}
\end{eqnarray}
with $\delta = \frac{\beta_1}{\beta_0} z_n$, i.e. $R(z)$ displays
a cut singularity \cite{Mue} at the renormalon position $z = z_n$, which 
generates 
according to Eq.~(3), since $z_n > 0$, an ${\cal O}(exp(-z_n/\alpha))$ imaginary
 part
in $R_{PT}(\alpha)$. In the standard case where $\alpha_{eff}$ has a Landau 
ghost 
at some scale $\Lambda^2 < Q^2$, this fact causes no surprise, since the 
defining 
integral Eq.~(1) itself involves an ambiguous integration over the Landau
singularity. But a paradox arises if $\alpha_{eff}$ has an IR fixed point : then
$R(\alpha)$ should be perfectly well defined, with no imaginary part according 
to 
Eq.~(1), whereas Eq.~(3), together with Eq.~(7), still yield an ambiguous, 
imaginary 
amplitude for $R_{PT}(\alpha)$ ! But Eq.~(7) is correct in both cases : as 
mentionned 
above, it follows solely from $RG$ invariance and the representation Eq.~(5) 
(which 
is presumably always valid since it is correct order by order in perturbation 
expansion in $z$, and the corresponding series are convergent). Thus $IR$ 
renormalons
{\em are also present in the IR fixed point case}. The only possible conclusion 
is that Eq.~(3) 
is not in fact a valid representation of Eq.~(1), i.e. that $R(\alpha)$ in 
Eq.~(1) does not
coincide with $R_{PT}(\alpha)$ in Eq.~(3) in the latter case. This is despite 
the fact I 
assumed Eq.~(4) does instead correctly represents (at least for large enough 
$k$) $\alpha_{eff}(k)$
(which is also assumed to have no renormalons), which can thus be determined in
 principle
from ``all order'' perturbation theory, and qualifies the present framework as 
being 
``perturbative''(the alternative possibility that $\alpha_{eff}(k)$
acquires its 
fixed point through additionnal ``higher-twist''type power like corrections not
 included in the Borel representation will be commented upon below). 

Indeed, there is a loophole in the previous derivation that $R(\alpha) = 
R_{PT}(\alpha)$,
which is actually already present when there is a Landau singularity. Assume 
e.g. $\alpha_{eff}$
is the one loop coupling :

\begin{eqnarray}
\alpha_{eff}(k/Q,\alpha) & = & {\displaystyle\frac {\alpha}{1 + \alpha\beta_0
\ ln \left(\frac{k^2}{Q^2}\right)}}
\nonumber\\
& \equiv & \frac{1}{\beta_0\ ln \left(\frac{k^2}{\Lambda^2}\right)}\nonumber
\end{eqnarray}

Then \cite{Par} : $\alpha_{eff}(k/Q,z) =  exp\lbrack -z \beta_0\ ln (k^2/Q^2)
\rbrack$ and Eq.~(5) yields :

$$
R(z) = \int_0^{Q^{2}} n \frac{dk^2}{k^2} \left(\frac{k^2}{Q^2}\right)^n
exp \left\lbrack -z\ \beta_0\ ln\ \left(\frac{k^2}{Q^2}\right)\right\rbrack 
= \frac{1}{1-\frac{z}{z_n}}
$$
The problem with the previous argument is that Eq.~(4) itself is a valid 
representation
of $\alpha_{eff}(k)$ {\em only if $k$ is not too small}. Indeed in the
present example the integral in Eq.~(4) converges at $z = \infty$
only if $\frac{1}{\alpha} + \beta_0\ ln \left(\frac{k^2}{Q^2}\right) > 0$,
i.e. for $k^2 > \Lambda^2$. For $k^2 < \Lambda^2$, $\alpha_{eff}(k)$
has to be represented instead by a Borel integral over the {\em negative}
$z$ axis :

\begin{eqnarray}
\alpha_{eff}(k/Q,\alpha) = - \int_{-\infty}^0 dz\ exp\ \left(-\frac{z}{\alpha}
\right)
\alpha_{eff} (k/Q,z)
\end{eqnarray}
Similar remarks apply in the general case, where Eq.~(6) implies, provided 
$\alpha_{eff}(z)$
decreases no faster then $exp(-cz)$ at large $z$,\ $\alpha_{eff}(k)$ will have 
a Landau singularity at some scale $k^2 = \Lambda^2$ (see however the fixed 
point 
case), below which the Borel representation Eq.~(4) will break down, and Eq.~(8)
should be used instead.

These observations suggest that the correct procedure in the Landau ghost case 
is to first 
take $\alpha < 0$ (and slightly complex if $\beta_1 \not= 0$), which implies 
$Q^2 < \Lambda^2$,
so that in the whole integration range in Eq.~(1) one has $k^2 < Q^2 < 
\Lambda^2$, and the 
representation Eq.~(8) is valid {\em throughout} the range (this means choosing
 $\alpha$ in the domain 
of attraction of the {\em trivial} $IR$ fixed point). Manipulations analoguous 
to those performed
above are now justified, and yield :

\begin{eqnarray}
R(\alpha) = - \int_{-\infty}^0 dz\ exp\ \left(- \frac {z}{\alpha}\right) R(z)
\end{eqnarray}
Finally, $R(\alpha)$ for $\alpha > 0$ (where $Q^2 > \Lambda^2$) is defined as 
the analytic
continuation of Eq.~(9), which yields the standard Borel representation Eq.~(3)
 with $R = R_{PT}$.

In the case where $\alpha_{eff}(k)$ has an IR fixed point, a similar problem 
arises if it happens
again that the representation Eq.~(4) is not valid below some scale $k = k_{min}
$, even if $k_{min}$
does not correspond to a Landau singularity this time. This is possible if, as 
in the Landau ghost case,
$\alpha_{eff}(z)$ does not decrease too fast at large $z$ (the latter assumption
 seems necessary, 
because a too fast decrease of $\alpha_{eff}(z)$ (e.g. \cite{Par} if $\alpha_
{eff}(z) = exp(-cz^2)$), although
insuring the validity of Eq.~(4) for all k's, usually yields an $\alpha_{eff}(k)
$ which blows up too fast 
as $k \rightarrow 0$, making the original integral Eq.~(1) IR divergent). 
Furthermore, {\em and this is the crucial
difference with the Landau ghost case}, for $k < k_{min}$\ ,  $\alpha_{eff}(k)$,
 which remains positive,
will not admit a Borel representation over the $z < 0$ axis either. (Note that 
for $k \rightarrow 0$,
$\alpha_{eff}(k)$ is always in the strong coupling region, since it approaches a
 non-trivial fixed point,
at the difference of the Landau ghost case, where for $k \rightarrow 0$ one 
enters an (IR) asymptotically
free region below the Landau singularity, since no other fixed point is 
available). The above derivation that
$R = R_{PT}$ can thus not be extended to the IR fixed point case, as expected.
 Let us now give a general
argument that these two functions actually differ by a power correction. 
Splitting the integration range
in Eq.~(1) at $k = k_{min}$, one has :

\begin{eqnarray}
R(\alpha) & = & \int_0^{k_{min}^2} n\ \frac{dk^2}{k^2} \left(\frac{k^2}{Q^2}
\right)^n
\alpha_{eff} \left(\frac{k}{Q},\alpha\right) + \int_{k_{min}^2}^{Q^2} n\ \frac
{dk^2}{k^2}
\left(\frac{k^2}{Q^2}\right)^n \alpha_{eff} \left(\frac{k}{Q},\alpha\right)
\nonumber\\
& \equiv &R_- + R_+ \nonumber
\end{eqnarray}
Eq.~(4) cannot be used inside the low momentum integral $R_-$, which can however
 be parametrized
as a power correction since :

\begin{eqnarray}
R_- & = & \left(\frac{k_{min}^2}{Q^2}\right)^n \int_0^{k_{min}^2} n\ \frac{dk^2}
{k^2}
\left(\frac{k^2}{k_{min}^2}\right)^n \alpha_{eff} \left(\frac{k}{k_{min}} , 
\alpha_{min}\right)\nonumber\\
& \equiv & \left(\frac{k_{min}^2}{Q^2}\right)^n R(\alpha_{min})\nonumber
\end{eqnarray}
where $\alpha_{min} = \alpha(Q = k_{min})$, and RG invariance has been used in 
the first step.
On the other hand, using Eq.~(4) in $R_+$, one gets :

$$
R_+ = \int_0^{\infty} dz\ exp \left(-\frac{z}{\alpha}\right) 
\int_{k_{min}^2}^{Q^2} n\ \frac{dk^2}{k^2} \left(\frac{k^2}{Q^2}\right)^n
\alpha_{eff} \left(\frac{k}{Q},z\right)
$$
To go further, it is necessary to know the k-dependence of $\alpha_{eff}(k/Q,z)$
. This can be
done easily in the special case where $\beta_1 = 0$, if one chooses $\alpha(Q)$
 to be the one 
loop coupling. Then one can show\cite{Par} that : $\alpha_{eff}(k/Q,z) = 
\alpha_{eff}(z) exp\lbrack -z \beta_0\ ln (k^2/Q^2)\rbrack$,
and one gets :

$$
\int_{k_{min}^2}^{Q^2} n\ \frac{dk^2}{k^2} \left(\frac{k^2}{Q^2}\right)^n
\alpha_{eff} \left(\frac{k}{Q},z\right) = \alpha_{eff}(z) \frac{1}{1-\frac{z}
{z_n}}
\left\lbrack 1 - \left(\frac{k_{min}^2}{Q^2}\right)^{z_n \beta_0 (1-z/z_n)}
\right\rbrack
$$
Hence :

\begin{eqnarray}
R_+ & = &\int_0^{\infty} dz\ exp \left(-\frac{z}{\alpha}\right) \alpha_{eff}(z)
 \frac{1}{1 - \frac{z}{z_n}}
- \left(\frac{k_{min}^2}{Q^2}\right)^n \int_0^{\infty} dz\ exp \left(- \frac{z}
{\alpha_{min}} \right)
\alpha_{eff}(z) \frac {1}{1 - \frac{z}{z_n}}\nonumber\\
& \equiv & R_{PT}(\alpha) - \left(\frac{k_{min}^2}{Q^2}\right)^n R_{PT}
(\alpha_{min})\nonumber 
\end{eqnarray}
One therefore ends up with the result :

\begin{eqnarray}
R(\alpha) = R_{PT}(\alpha) + \left(\frac{k_{min}^2}{Q^2}\right)^n \lbrack 
R(\alpha_{min}) - R_{PT}(\alpha_{min})\rbrack
\end{eqnarray}
where the coefficient of the power correction is given by the discrepancy 
between the exact
amplitude and its Borel representation. Eq.~(10) is equivalent to the statement 
that 
$R(\alpha) = R_{PT}(\alpha) + const/(Q^2)^n$, and is also correct when 
$\beta_1 \not= 0$ (see below). 
Note it is not possible the power correction vanishes (as it does in the Landau
 ghost case)
since $R(\alpha_{min})$ is real, whereas $R_{PT}(\alpha_{min})$ is complex due 
to the effect of the renormalon (the power correction must be 
complex to cancell the imaginary part of $R_{PT}(\alpha))$.

I now illustrate the previous discussion with the example of the 2 loop coupling
 :

\begin{eqnarray}
\frac{d\alpha_{eff}}{d ln k^2} = - \beta_0 (\alpha_{eff})^2 - \beta_1 (
\alpha_{eff})^3
\end{eqnarray}
I shall consider both the standard case $\beta_1/\beta_0 > 0$ where there is a
 Landau singularity,
and the case $\beta_1/\beta_0 < 0$ where $\alpha_{eff}(k)$ has an IR fixed point
 at $\alpha_{IR}  -\beta_0/\beta_1$ (which actually occurs in QCD for a large enough number of 
flavors). Remarkably, 
$R(z)$ can be computed {\em exactly} with a straightforward change of variable, 
adapted from a similar one suggested in ref.3.
Defining the Borel variable by :

\begin{eqnarray}
\frac{z}{z_n} = \frac{1 - \frac{\alpha}{\alpha_{eff}(k)}}{1 + \frac{\beta_1}
{\beta_0} \alpha}
\end{eqnarray}
with $\alpha \equiv \alpha_{eff} (k = Q)$, and using the solution of Eq.~(11) :

\begin{eqnarray}
ln \left(\frac{k^2}{\Lambda^2}\right) = \frac {1}{\beta_0 \alpha_{eff}} - 
\frac{\beta_1}{\beta_0^2}
\ ln \left(\frac{1}{\alpha_{eff}} + \frac{\beta_1}{\beta_0}\right) + \frac
{\beta_1}{\beta_0^2}
\end{eqnarray}
(together with the similar relation at $k = Q$) one obtains :

$$
n \frac{dk^2}{k^2} \left(\frac{k^2}{Q^2}\right)^n \alpha_{eff}(k) = - dz 
\frac{exp (-\frac{z}{\alpha}
- \frac{\beta_1}{\beta_0} z)}{\left(1 - \frac{z}{z_n}\right)^{1 + \delta}}
$$
which suggests the looked for Borel transform is :

\begin{eqnarray}
R(z) = \frac{exp \left(- \frac{\beta_1}{\beta_0} z\right)}{\left(1 - \frac{z}
{z_n}\right)^{1+\delta}}\ .
\end{eqnarray}
To complete the proof, it remains to determine the new integration bounds.
\vskip 1truecm
1) Assume first $\beta_1 / \beta_0 > 0$ : then $\alpha_{eff}$ has a Landau 
singularity
at the scale ${\bar\Lambda}^2 = \Lambda^2 exp \left(- \frac{\beta_1}{\beta_0^2} 
ln \left(\frac{\beta_1}{\beta_0}\right) + \frac{\beta_1}{\beta_0^2}\right)$.
I assume $Q^2 > {\bar\Lambda^2}$, so that $\alpha > 0$. For $k^2 = Q^2$,
 $\alpha_{eff} = \alpha$,
and $z = 0$ ; at $k^2 = {\bar\Lambda^2}$ , $\alpha_{eff} = \infty$, and 
$z = z_L \equiv 
\frac{z_n}{1 + \frac{\beta_1}{\beta_0} \alpha} < z_n$ ; decreasing $k^2$ below 
${\bar\Lambda}^2$,
$z$ becomes complex ; finally, for $k^2 \rightarrow 0$, $\alpha_{eff} 
\rightarrow 0^-$, and
$z \rightarrow +\infty$. $R(\alpha)$ in Eq.~(1) then takes the form of a Borel 
integral along a 
path in the complex $z$-plane, which divides into two complex conjuguate 
branches at $z = z_L$, 
and can be deformed to above or below the positive real axis to yield the 
standard Borel representation
Eq.~(3) with $R(z)$ as in Eq.~(14).
\vskip 1truecm
2) On the other hand, if $\beta_1/\beta_0 < 0$, the situation is actually 
simpler : as $k^2$ decreases
from $Q^2$ to $0$, $\alpha_{eff}$ increases from $\alpha$ to the IR fixed point
 $\alpha_{IR}$, and $z$
increases from $0$ to $z_n$ through real values so that Eq.~(1) becomes :

\begin{eqnarray}
R(\alpha) = \int_0^{z_n} dz\ exp\ \left(- \frac{z}{\alpha}\right) 
\frac{exp\left(-\frac{\beta_1}{\beta_0}\ z\right)}{\left(1 - \frac{z}{z_n}
\right)^{1+\delta}}
\end{eqnarray}
(the renormalon singularity is integrable at $z = z_n$, since $\delta < 0$ now).
 We therefore
check that $R(\alpha)$ is {\em not} given by the Borel sum $R_{PT}(\alpha)$ in 
the IR fixed point case.
Rather, it is given by $R_{PT}(\alpha)$ minus an {\em exponentially small} 
${\cal O}(exp(-z_n/\alpha))$ correction~:

\begin{eqnarray}
R(\alpha) & = & \int_0^{\infty} dz\ exp\ \left(- \frac{z}{\alpha}\right) 
\frac{exp\left(-\frac{\beta_1}{\beta_0}\ z\right)}{\left(1 - \frac{z}{z_n}
\right)^{1+\delta}}
- \int_{z_n}^{\infty} dz\ exp\ \left(- \frac{z}{\alpha}\right) 
\frac{exp\left(-\frac{\beta_1}{\beta_0}\ z\right)}{\left(1 - \frac{z}{z_n}
\right)^{1+\delta}}\nonumber\\
& \equiv & R_{PT} + R_{NP}
\end{eqnarray}
The latter is just a (complex) power correction, in accordance with the general
 \\expectation.
Indeed, Eq.~(15)-(16) simplify by performing the change of coupling : 
$1/a = 1/\alpha + \beta_1/\beta_0$,
and one gets in particular :\\

\begin{eqnarray}
R_{NP} = - \int_{z_n}^{\infty} dz\ exp\ \left(-\frac{z}{a} \right) 
\frac{1}{\left(1 - \frac{z}{z_n}\right)^{1+\delta}} 
& = & - \tilde{C}\ exp\ (-z_n/a)(-1/a)^{\delta}\nonumber\\
& = & - \tilde{C}(-1)^{\delta}\ (\Lambda^2/Q^2)^n
\end{eqnarray}
with $\tilde{C} = \frac{\beta_0}{\beta_1} \Gamma(1-\delta)(z_n)^{\delta}$, and
 Eq.~(13) at $k = Q$
was used in the last step (in this example , the subtracted term $-R_{NP}$ 
corresponds exactly
to the ``minimal'' prescription of ref.4 to ``regularize'' IR renormalons).
The same method can deal with the case $\alpha < 0$, where one is in the domain
 of attraction
of the {\em trivial} IR fixed point (provided the condition $1 + \frac{\beta_1}
{\beta_0} \alpha > 0$
is also satisfied if $\beta_1/\beta_0 < 0$). Then $\alpha_{eff}(k)$ monotonously
 increases from $\alpha$
to $0^-$ as $k^2$ decreases from $Q^2$ to 0, hence $z < 0$ and decreases from 0
 to $-\infty$. Thus :

\begin{eqnarray}
R(\alpha) = -\int_{-\infty}^0 dz\ exp\ \left(-\frac{z}{\alpha}\right) \frac
{exp\left(-\frac{\beta_1}{\beta_0} z\right)}
{\left(1 - \frac{z}{z_n}\right)^{1+\delta}}
\end{eqnarray}
i.e. a Borel integral over the {\em negative} z axis, in agreement with Eq.~(9)
 (note that Eq.~(15) is
{\em not} the analytic continuation of Eq.~(18)).

For completness I give also the result if $\alpha_{eff}(k)$ satisfies the RG
 equation :

$$
\frac{d\alpha_{eff}}{d\ ln\ k^2} = \frac{-\beta_0\ \alpha_{eff}^2}{1 - \frac
{\beta_1}{\beta_0} \alpha_{eff}}
$$
where the {\em inverse} beta function has only two terms. The appropriate change
 of variable in this case turns out
to be precisely the one suggested in ref.3 :

$$
\frac{z}{z_n} = 1 - \frac{\alpha}{\alpha_{eff}(k)}
$$
(with $\alpha \equiv \alpha_{eff}(k = Q))$, which yields for $\beta_1/\beta_0 >
 0$
(using also the convolution theorem) :

$$
R(\alpha) = \frac{\delta}{1 + \delta}\ \alpha + \frac{1}{1 + \delta} \int_0^
{\infty}
dz\ exp\ \left(-\frac{z}{\alpha}\right) \frac{1}{\left(1 - \frac{z}{z_n}
\right)^{1+\delta}}
$$
whereas for $\beta_1/\beta_0 < 0$, where $\alpha_{eff}$ has an {\em infinite} 
IR
 fixed point,
one gets :

$$
R(\alpha) = \frac{\delta}{1+\delta}\ \alpha + \frac{1}{1+\delta} 
\int_0^{z_n} dz\ exp\ \left(-\frac{z}{\alpha}\right)
\frac{1}{\left(1 - \frac{z}{z_n}\right)^{1+\delta}}
$$
and shows that in this case too (with a different $\tilde{C}$) :
$R(\alpha) = R_{PT}(\alpha) - \tilde{C}(-1)^{\delta}
(\Lambda^2/Q^2)^n$.

For an arbitrary $\beta_{eff}$ function with an IR fixed point one 
expects however the answer to be of the more general form :

\begin{eqnarray}
R(\alpha) = R_{PT}(\alpha) + \left(\frac{\Lambda^2}{Q^2}\right)^n 
(-\tilde{C}(-1)^{\delta} + C)
\end{eqnarray}
where $C$ and $\tilde{C}$ are real, and independent of $Q$, and $\tilde{C}$
is proportionnal to the renormalon residue. It may be worth mentionning that 
taking the 
$Q^2$ derivative of both sides of Eq.~(19), and going to Borel space, one easily
 derives 
a relation between $R(z)$ and the Borel transform of $\beta_{eff}(\alpha) = 
d\alpha/dln\ Q^2$
(with $\alpha \equiv \alpha_{eff} (k = Q))$ :

\begin{eqnarray}
R(z) - \frac{\beta_0}{n} z\ R(z) - \frac{1}{n} \int_0^z dy\ b(z - y) y\ R(y) = 1
\end{eqnarray}
where $b(z)$ is the Borel transform  of $b(\alpha) \equiv - \frac{\beta_{eff}
(\alpha)}{\alpha^2} - \beta_0$.
Eq.~(20) can be used to rederive the previous two results, and allows to deal
 with more 
complicated examples as well. Note also that $C$ and $\tilde{C}$, being 
independent of $Q$,
have dropped from Eq.~(20), and in fact any $Q$-dependence in these coefficients
 would be
inconsistent with the assumption of the perturbative nature of the coupling 
(equivalent
here to assuming that $\beta_{eff}(\alpha)$ coincides with its Borel sum).

\underline{Non perturbative IR fixed point}: in the Landau ghost case,one has to
 add\\ explicitly power suppressed non perturbative terms $\alpha_{NP}$ to
 remove
the Landau ghost and generate an IR fixed point.Putting $\alpha_{eff}\alpha_{PT}+\alpha_{NP}$ ,where $\alpha_{PT}$ is assumed to have a Landau 
singularity and to satisfy the Borel representation Eq.~(4) we thus have:$R(Q)R_{PT}(Q)+R_{NP}(Q)$ where $R_{PT}(Q)$ is given by Eq.~(3) ,and:
\begin{eqnarray}R_{NP}(Q)&=&\int_{0}^{Q^{2}} n \frac{dk^{2}}{k^{2}} 
\left(\frac{k^{2}}{Q^{2}}\right)^{n}\alpha_{NP}(k)\nonumber\\
                         &=&\left(\frac{\Lambda^2}{Q^2}\right)^n 
[-\tilde{C}(-1)^{\delta} + C(Q)]\end{eqnarray}
where $C(Q)$ is real and represents the part of the power corrections unrelated
to renormalons.To make contact with ref.5 ,I now assume the integral in
Eq.~(21) converges at infinity ,in which case $C(Q)\rightarrow C$ at large $Q$ ,
with :
\begin{equation}-\tilde{C}(-1)^{\delta} + C = \int_{0}^{\infty} n 
\frac{dk^{2}}{k^{2}} \left(\frac{k^{2}}{\Lambda^{2}}\right)^{n}\alpha_{NP}(k)
\end{equation}
and,neglecting higher order power corrections :
\begin{equation}R(Q)\simeq R_{PT}(Q)+\left(\frac{\Lambda^2}{Q^2}\right)^n 
(-\tilde{C}(-1)^{\delta} + C)\end{equation}
Assume now the high energy approximation Eq.~(23) is valid for $Q\gsim\mu_I$ .
This is equivalent to assume that the high energy tail of the integral in 
Eq.~(22) can be neglected for $k\gsim\mu_I$ .Then from Eq.~(23) and its 
counterpart
at $Q=\mu_I$ one gets immediately:
\begin{equation}R(Q)\simeq R_{PT}(Q) + \left(\frac{\mu_I^2}{Q^2}\right)^n 
\lbrack R(\mu_I) - R_{PT}(\mu_I)\rbrack\end{equation}
which is the parametrisation of the power corrections in term of low energy 
integrals over the
full (IR regular) and the perturbative effective couplings suggested in ref.5.
In 
this approach,the renormalon problem is completely bypassed,since Eq.~(24) can 
be
equivalently written as:
\begin{equation}R(Q)\simeq\int_{\mu_I^2}^{Q^2} n\ \frac{dk^2}{k^2}
\left(\frac{k^2}{Q^2}\right)^n \alpha_{PT} \left(\frac{k}{Q},\alpha\right)+
\left(\frac{\mu_I^2}{Q^2}\right)^n  R(\mu_I)\end{equation}
and the integral over $\alpha_{PT}$ , being cutoff in the infrared,has no 
renormalons.
\\Perturbation theory,if used to estimate the integral in Eq.~(25) , is then
expected to be convergent at arbitrary high order.Eq.~(25) amounts to neglect
$\alpha_{NP}(k)$ for $k\gsim\mu_I$~, which is the assumption of ref.5.We have 
seen
in the example of the integral of Eq.~(1) this assumption can be justified if
$\alpha_{NP}(k)$ decreases sufficiently fast at large $k$~. Furthermore we have
 shown that
Eq.~(23) and (24) are actually {\em exact} (Eq.~(10) and (19)) in the case where
$\alpha_{NP}=0$ and the fixed point is of perturbative origin (then Eq.~(25) is
trivial),provided one interprets $R_{PT}(Q)$ as the Borel sum of perturbation 
theory.

\vskip 1truecm
\noindent
{\bf Acknowledgements}

I thank G. Marchesini and A. Mueller for useful discussions.

\end{document}